\documentclass{sig-alternate}

\usepackage{graphicx}
\usepackage{epstopdf}
\usepackage{caption}
\usepackage{subcaption}
\usepackage{lipsum,multicol}
\usepackage{hyphenat}
\begin{document}

\title{On the energy efficiency of client-centric data consistency
  management under random read/write access to Big Data with Apache
  HBase}

\numberofauthors{1}

\author{ \alignauthor
  \'Alvaro Garc\'ia-Recuero\\
  \affaddr{Inria Rennes - Bretagne Atlantique}\\
  \affaddr{Rennes, France}\\
  \email{alvaro.garcia-recuero@inria.fr}} \date{27 November 2014}

\maketitle
\begin{abstract}
  The total estimated energy bill for data centers in 2010 was \$11.5
  billion, and experts estimate that the energy cost of a typical data
  center doubles every five years. On the other hand, computational
  developments have started to lag behind storage advancements,
  therein becoming a future bottleneck for the ongoing data growth which
  already approaches Exascale levels. We investigate the relationship
  among data throughput and energy footprint on a large storage
  cluster, with the goal of formalizing it as a metric that reflects
  the trading among consistency and energy. Employing a client-centric
  consistency approach, and while honouring ACID properties of the
  chosen columnar store for the case study, Apache HBase, we present
  the factors involved in the energy consumption of the system as well
  as lessons learned to underpin further design of energy-efficient
  cluster scale storage systems. We only show most releavant
  preliminary results.
\end{abstract}

\category{H.3.4}{Systems and Software}{Performance
  evaluation}[efficiency and effectiveness]
\keywords{Energy-efficiency, Energy metrics, Big Data}

\section{Introduction}
As Big Data approaches Exascale levels, storage systems start to
experiment new challenges in regards to the volume, variety and speed
at which information needs to be processed (velocity). Energy costs
are a raising concern, as servers are not designed to be
power-proportional~\cite{Barroso:2007} and modern networks eventually
may not be able to cope with an already oversubscribed model beyond
access routers. In particular, the power dis-proportionality of
storage systems is usually due to the heterogeneous consumption of
disks as well as memory instability~\cite{Guerra:2010}.

In database applications,~\cite{Tsirogiannis:2010} explored the
energy-efficiency of databases, and found they are not able to measure
noticeable variations in power consumption using different workloads
when varied the amount of memory accessed and the access patterns
applied (sequential vs random memory accesses).

On the other hand, and while shared-nothing architectures allow
decoupling of the underlying hardware infrastructure from the
computation, it is yet still not fully understood how to efficiently
adapt distributed storage systems that sustain high throughput to
mission critical application's and reduce overall energy bills.

\section{Experimental study}
We study HBase, a column-oriented data store which follows the
architectural design of BigTable, and it is suited for random,
real-time read/write access to Big Data. Our findings show potential
for improvement in this research area. To the best of our knowledge,
this is the first study to date which focuses on the energy footprint
of random read and write workloads in a modern NoSQL data store (e.g.,
Apache HBase). To this end, we show the impact different workload
types, consistency and concurrency levels have over the total energy
consumption of the storage cluster as well as its data
throughput. Empirical results are obtained through automated and
reproducible experiments developed for running an HBase cluster of
machines on the Grid5000~\cite{grid5000} platform.

\subsection{Approach}
Our methodology follows a client-centric consistency model with two
configurations. Deferred-updates, with a buffer of size 12MB (default
in HBase), namely \emph{eventual}; or without buffer, namely
\emph{strong}. Both leverage the default Hadoop packet size of 64KB
(which in turn involves no buffer-copy). Naturally, HBase provides
strong consistency semantics at the row level and within a data
center. Therefore, for analyzing the effect of deferring or not
updates under a strongly consistent architecture, we embed these
semantics into the HBase client of YCSB (\emph{Yahoo Cloud Service
  Benchmark}). At the time of running the experiments we used a stable
version of HBase (hbase-0.94.8) in a cluster of 40 server machines of
the model Carri System CS-5393B with Intel Xeon X3440 CPU at 2.53 Ghz,
16 GB memory, 320GB / SATA II (drive ahci) of storage and Gigabit
Ethernet network connectivity. We experiment with 3 large workloads
that fully stress memory and therefore exercise hard-disks. Namely,
write intensive (80\% writes) as in e-Commerce applications during
peak season due to flash-crowds (\emph{e.g., Black Friday or Cyber
  Monday}), read intensive (80\% reads) which is the usual pattern in
HBase with the \emph{messages} application at Facebook, and balanced
in order to see the effect of a mixed workload (50\%-50\%). All of
them use a uniform data distribution in order to simulate random
reads/writes to HBase, meaning choosing an item uniformly at
random. Energy measurements are obtained through an API connected to
the \emph{power distribution units} in the data center.

\subsection{Modeling the trade-off}
With energy efficiency generally described as
$\frac{{Work Done}}{Energy Used}$, we realize the impact of
deferred-updates (as in eventually consistent systems) as the fraction
of throughput produced with a given amount of energy to be consumed by
the cluster. The hypothesis is that consistency guarantees (latency of
update propagation time) are offered to the client in exchange of a
given energy footprint on the cluster. A simple model can introduce a
good estimation of this trade-off by considering the different factors
leading to a final amount of energy consumed; conveyed as the amount
of consistency offered with a given energy budget in fact,
$\frac{{Consistency SLA}}{Energy Budget}$.

\begin{figure}[!b]
  \begin{minipage}{\textwidth}
    \centering
    \begin{tabular}{ccc}
      \includegraphics[width=0.25\textwidth]{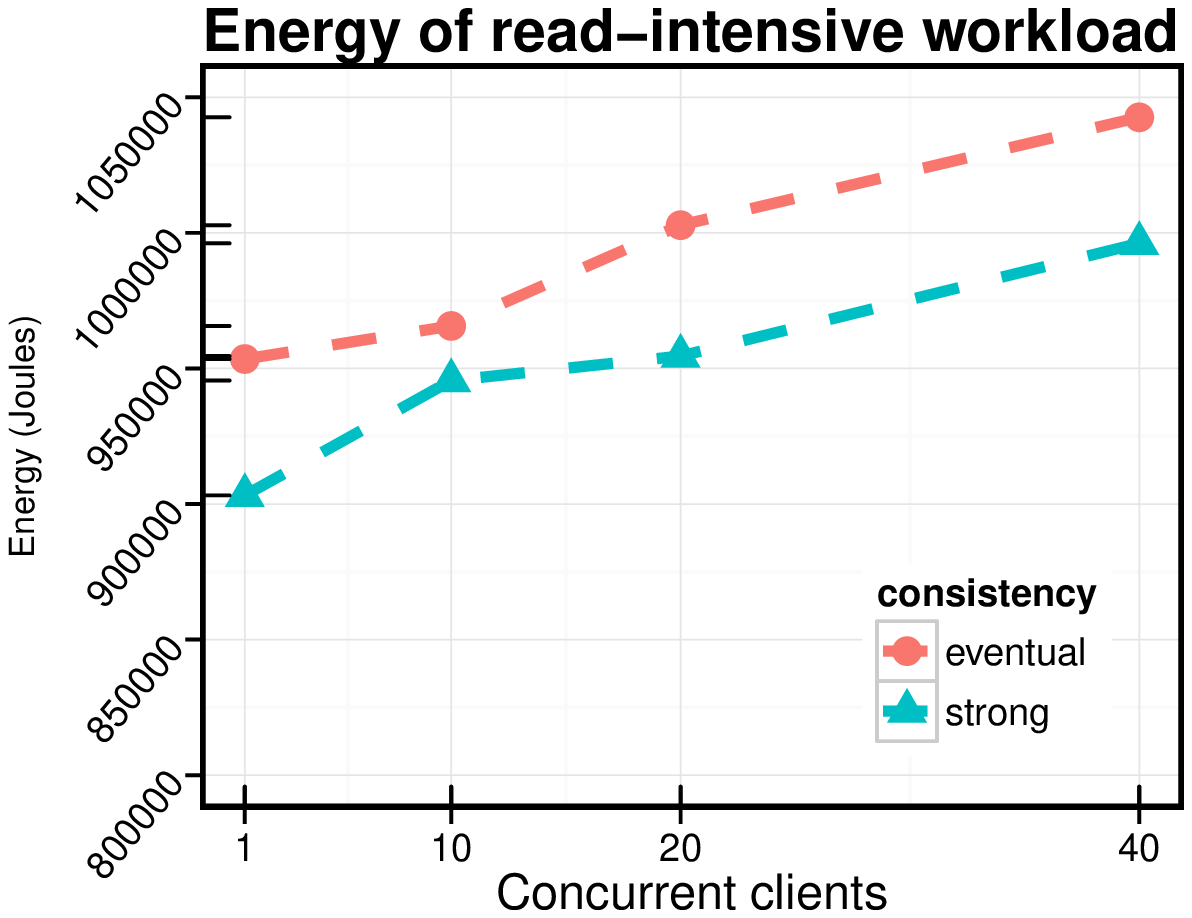}\label{fig:readsE}
      & 
        \includegraphics[width=0.25\textwidth]{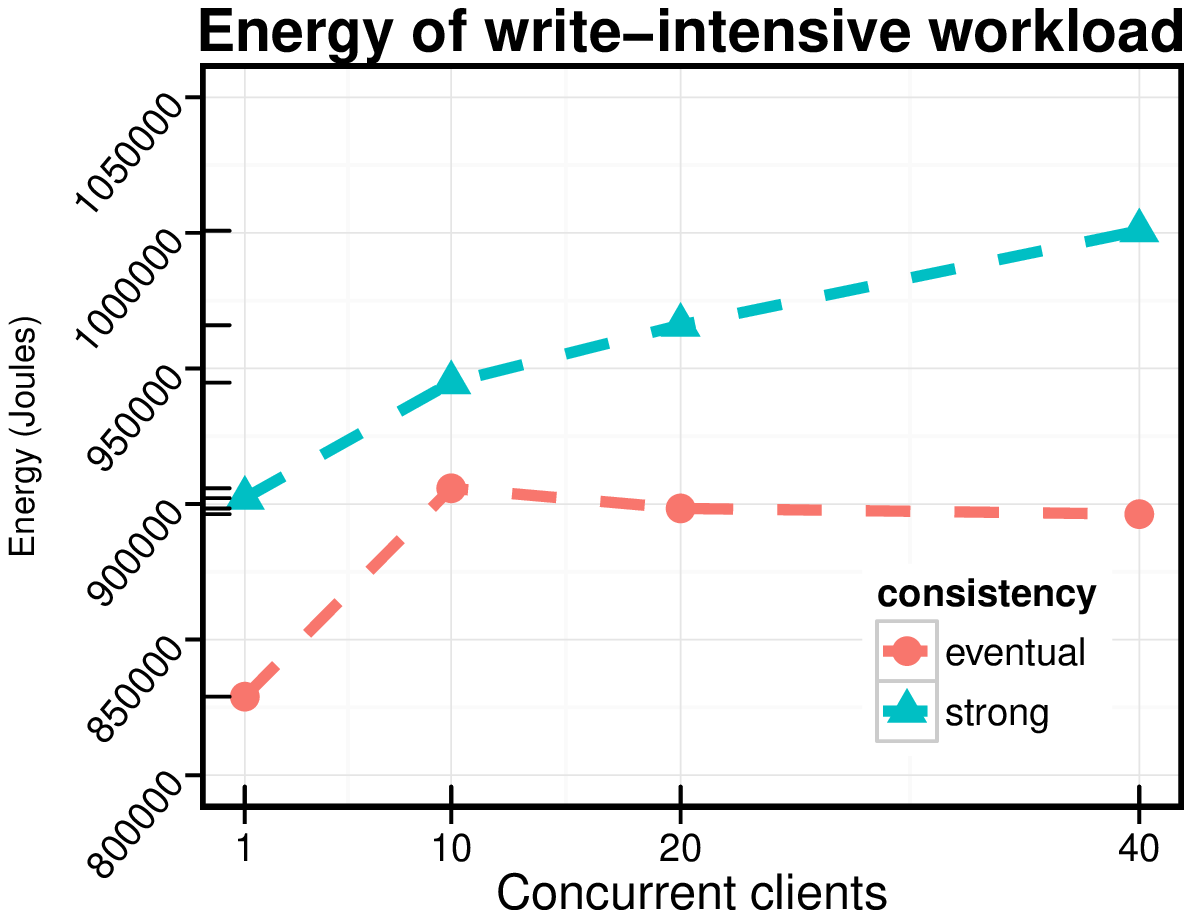}\label{fig:writesE}
      &  \includegraphics[width=0.25\textwidth]{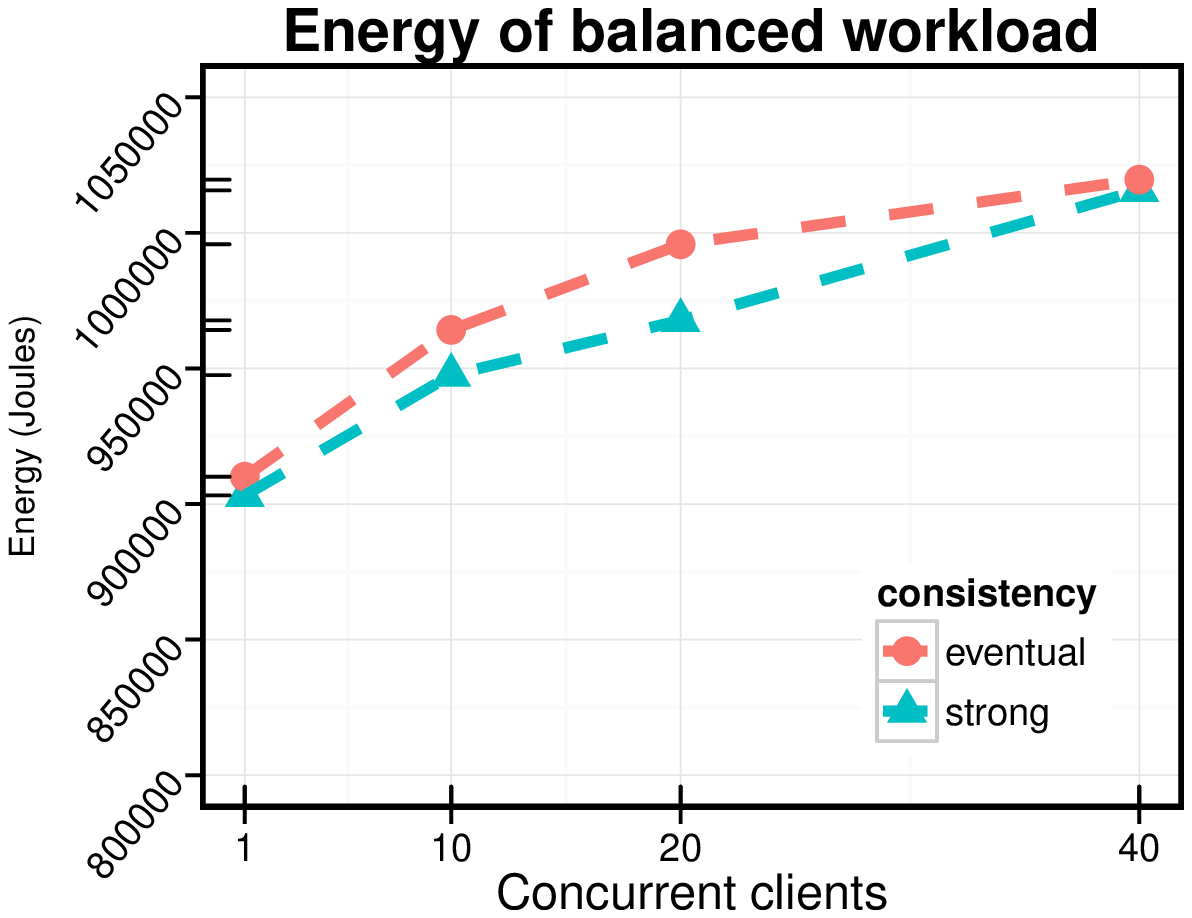}\label{fig:balancedE}
      \\
      (a) energy reads & (b) energy writes & (c) energy balanced \\[6pt]
      \includegraphics[width=0.25\textwidth]{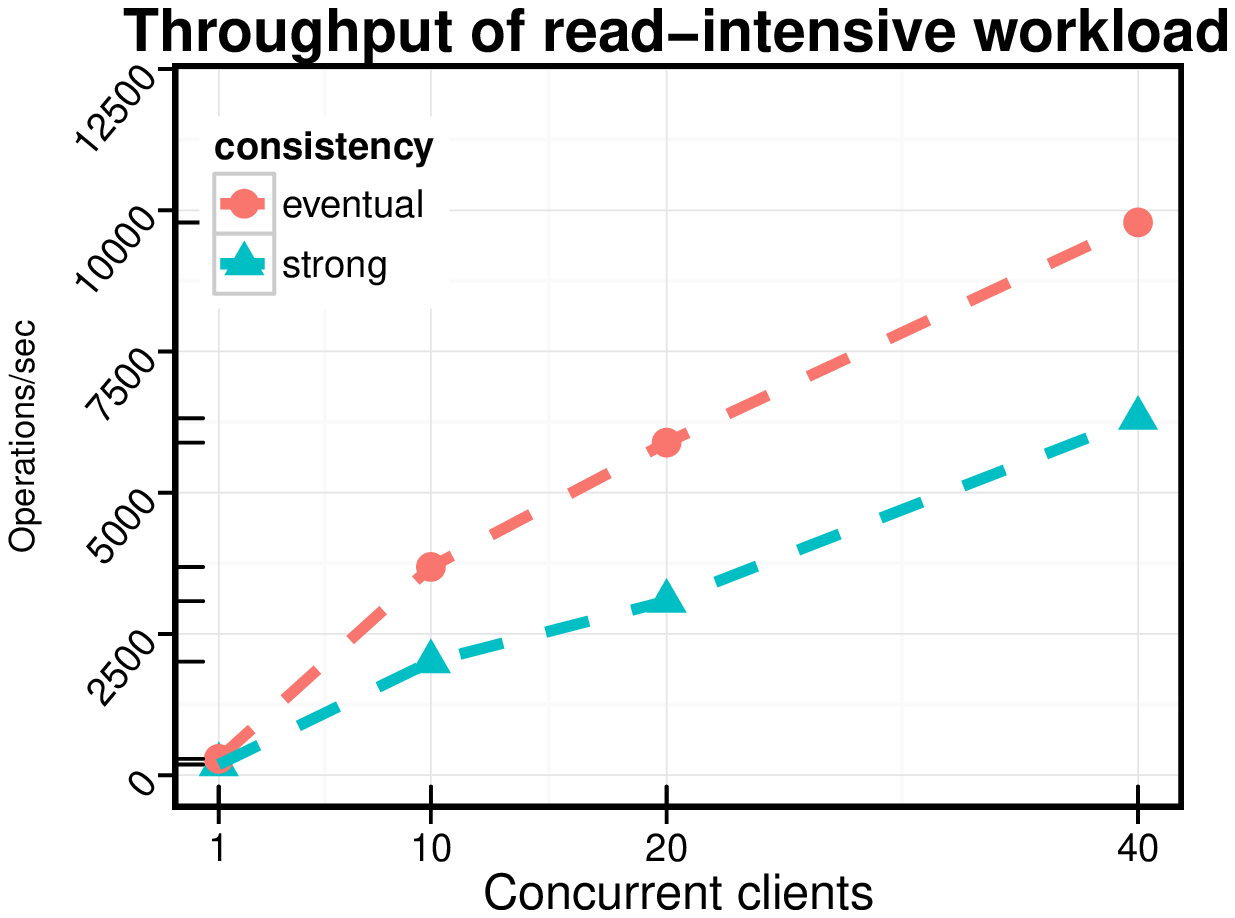}\label{fig:readsT}
      & \includegraphics[width=0.25\textwidth]{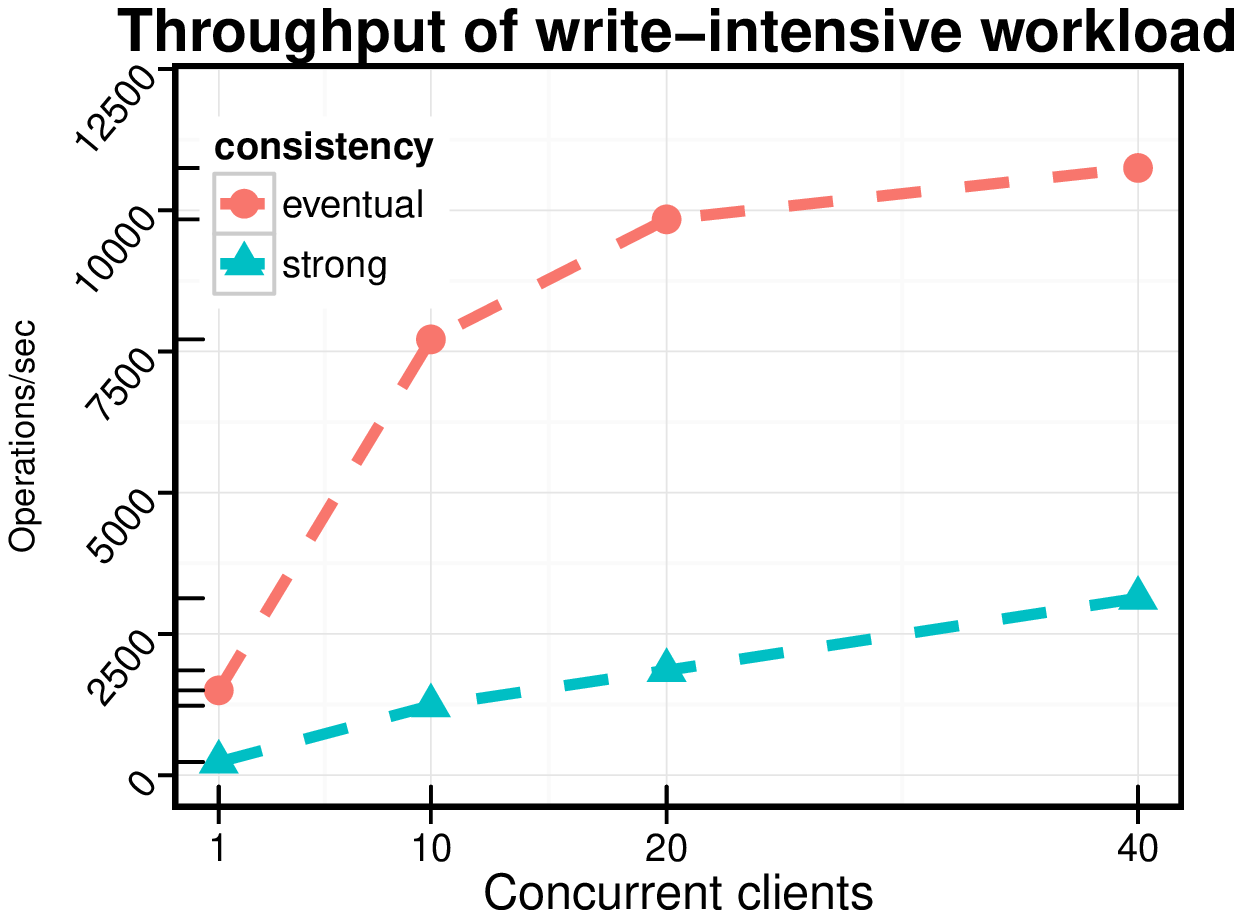}\label{fig:writesT}
      & \includegraphics[width=0.25\textwidth]{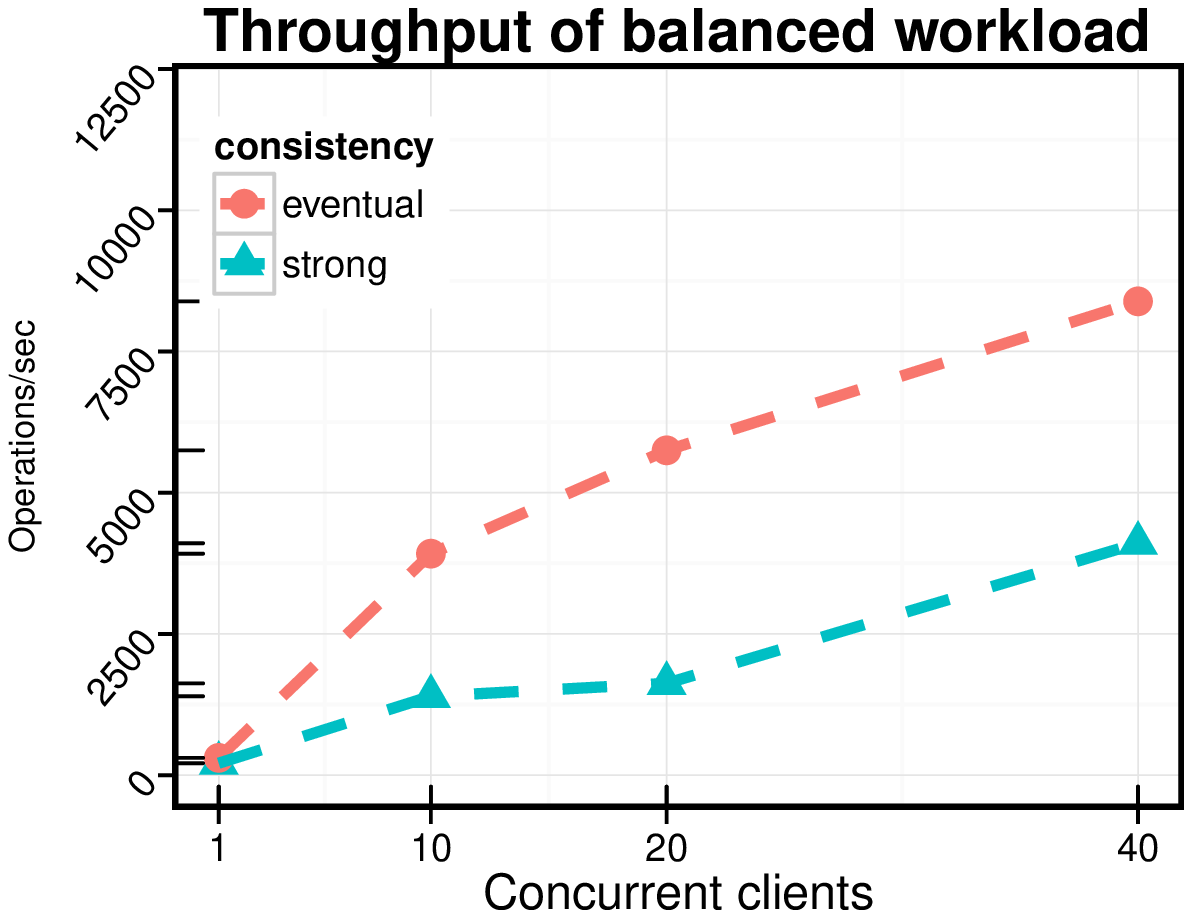}\label{fig:balancedT} \\
      (d) throughput reads & (e) throughput writes & (f) throughput balanced \\[6pt]
    \end{tabular}
    \caption{Energy Vs Speedup}
  \end{minipage}
\end{figure}

\section{Analysis and conclusions}
In this paper we analyze and characterize the energy efficiency of
three different workloads, which exhibit different behaviors in terms
of energy consumption and data throughput. While strong delivers
poorer performance and often consumes same or more energy than
eventual (as in the case of a write intensive workload under high
concurrency), there is a substantial improvement in throughput when
using eventual in all cases.

The most interesting case is the write workload. \emph{Eventual} (with
buffer) achieves around 3x times higher throughput under high
concurrency and averages about 0,01 Kilowatt-hour (kW*h) less than the
\emph{strong} approach (without buffer). Those savings increase as the
number of concurrent clients grow because of the steady consumption
with \emph{strong}, unlike the case of \emph{eventual}. The case of
reads is more surprising, which reveals that in systems such as HBase,
built on top of a memory store, reads cost more energy per unit of
throughput. The balanced workload follows the same trend as well,
indicating the clear impact of reads once again.

Therefore, access patterns, concurrency and consistency leads to a
given energy consumption for each type of workload. This is
highlighted as the relationship among \emph{Energy} and
\emph{Throughput} in a modern data store that is built to scale with
random reads and writes.

In turn, it must be possible to reach further energy savings by
applying~\emph{Write off-loading} techniques on HBase idle region
servers pointing to a distributed and common file system (HDFS). That
is, changing requests patterns by caching or moving such requests from
the unused disks into another location in the data center, and
therefore expecting to increase energy savings substantially as
in~\cite{Narayanan:2008}.

\section{Acknowledgments}
Experiments presented in this paper were carried out using the
Grid'5000 experimental testbed, being developed under the INRIA
ALADDIN development action with support from CNRS, RENATER and several
Universities as well as other funding bodies (see
https://www.grid5000.fr).

\small{ \bibliographystyle{abbrv}
  \bibliography{sigproc} 
}
\end{document}